\documentclass[useAMS,usenatbib]{mn2e}
\usepackage{graphicx}
\usepackage{longtable}
\title[The He~{\sc i} outflow in NGC 4151]{Physical properties and the variability mechanism of the He~{\sc i} outflow in NGC 4151}
\newcommand*\samethanks[1][\value{footnote}]{\footnotemark[#1]}
\author[C. Wildy et al.]{C. Wildy$^{1}$\thanks{E-mail:
    cw268@le.ac.uk}, H. Landt$^{2}$\thanks{Visiting Astronomer at the
    Infrared Telescope Facility, which is operated by the University
    of Hawaii under Cooperative Agreement no. NNX-08AE38A with the
    National Aeronautics and Sp ace Administration, Science Mission
    Directorate, Planetary Astronomy Program.}, M. R. Goad$^{1}$, M. Ward$^{2}$\samethanks{} and J. S. Collinson$^{2}$\\ 
$^{1}$University of Leicester, Department of Physics and Astronomy, University Road, Leicester, UK\\
$^{2}$Department of Physics, Durham University, South Road, Durham DH1 3LE, UK}

  \renewenvironment{thebibliography}[1]{%
    \begin{oldthebibliography}{#1}%
      \setlength{\parskip}{0ex}%
      \setlength{\itemsep}{0ex}%
  }%
  {%
    \end{oldthebibliography}%
  }

\begin{document}

\date{Received xxx; in original form
  January 0000}

\pagerange{\pageref{firstpage}--\pageref{lastpage}} \pubyear{0000}

\maketitle

\label{firstpage}

\begin{abstract}

\noindent We report on variable helium absorption lines in NGC~4151 observed across six epochs of quasi-simultaneous near-infrared and optical data. These observations cover the transitions from the metastable $2^{3}S$ state at 3889~\AA{} and 10\,830~\AA{}, and from the $2^{1}S$ state at 20\,587 \AA{}. This is the first AGN absorption line variability study to include measurements of the 20\,587~\AA{} line. The physical properties of the absorber recorded at the fifth observational epoch are relatively well constrained by the presence of absorption in both the optical and near-infrared components, with the 10\,830~\AA{} line likely saturated. The observations suggest variations in this absorber's strength are best explained by ionization changes in response to a variable incident continuum. Photoionization simulations constrain the total hydrogen number density of the epoch~5 absorber to 7.1$\leq$log($n_{{\rm H}}$/cm$^{-3}$)$\leq$8.8, the hydrogen column density to 21.2$\leq$log($N_{{\rm H}}$/cm$^{-2}$)$\leq$23.3 and the ionization parameter range to $-$1.9$\leq$log$U$$\leq$0.4. The simulations also suggest the absorber is located between 0.03 and 0.49~pc from the continuum emission region. This range in physical properties is consistent with an absorber of similar velocity seen in NGC~4151 from previous ultraviolet and optical studies, but with high column density X-ray absorbing components not present. The mass outflow rate due to the fifth epoch absorber is in the range 0.008 to 0.38~$M_{\odot}$~yr$^{-1}$, too low to contribute to galaxy feedback effects.

\end{abstract}


\section{Introduction}

Outflows from Active Galactic Nuclei (AGN) can be manifested by blueshifted absorption lines and are detected in wavelength bands ranging from X-rays to the near-infrared (hereafter, near-IR). These outflows can attain velocities of up to 20 per cent of the vacuum light speed in the host galaxy rest-frame \citep{pounds03}, and may be accelerated by radiation pressure, magnetic effects or thermal expansion \citep{chelouche01,everett05,giustini12}. Outflows are thought to have a significant impact on the host galaxy and may regulate black growth via feedback \citep{magorrian98,silk98,scannapieco04,wang06,nardini15}. Many absorption features in AGN are observed to undergo variability across a wide range of timescales spanning hours to several years \citep{guido10,filizak13}. The mechanism responsible for this variability is suggested to be either movement of gas across the line-of-sight or changes in its ionization state \citep{arav98,hamann08}.

The Seyfert~1 Galaxy NGC~4151 has undergone intensive study across the electromagnetic spectrum since it hosts one of the closest and brightest active nuclei as seen from Earth. One of its most intriguing characteristics is the dramatic variability in the strength of the continuum emission, a phenomenon which is observed in X-ray, ultraviolet (UV) and optical wavebands \citep{edelson96}. The spectral variability of the absorption lines has, however, been less intensively studied. An investigation of the narrow line region (NLR) in NGC~4151 has revealed a possible geometry for the outflowing gas, indicating that it originates in a conical region with a full opening angle of $\sim$80$^{\circ}$ with a line-of-sight viewing angle lying 10$^{\circ}$ outside of the cone \citep{hutchings98}.

Optical spectroscopic observations obtained using the Space Telescope Imaging Spectrograph (STIS) aboard the Hubble Space Telescope (HST) indicated that Balmer and metastable helium absorption features, which had previously only been observed during low continuum states, were also present during high continuum states. They also measured higher outflow velocities at higher continuum fluxes \citep{hutchings02}. A series of papers \citep{crenshaw00,kraemer01,kraemer05,kraemer06,crenshaw07} reported that the absorption features seen in the UV and optical transitions across a range of ionization states were dominated by a feature labelled ``D+E''. This feature had a radial velocity centroid of $\approx$$-$500~km~s$^{-1}$ relative to the rest-frame of the source with a full-width at half-maximum (FWHM) of 1170~km~s$^{-1}$. The absorbing material was found to be located at a distance of $\sim$0.1~pc from the central engine with a number density of between 10$^{7}$ and 10$^{9}$~cm$^{-3}$.

A situation frequently encountered when analysing AGN absorption lines is that of saturation at non-zero residual intensities. A signature of this effect is when two components of an absorber have an observed ratio of line optical depths closer to 1:1 than that which atomic theory predicts, implying the stronger component has reached opacity over at least part of its wavelength span. This is especially noticeable in many UV absorbers having doublet structures such as Si~{\sc iv} $\lambda$$\lambda$1394,1403 and C~{\sc iv} $\lambda$$\lambda$1548,1551 where the predicted optical depth ratio is 2:1. In these cases the effect is interpreted as resulting from an absorber which only partially covers the emission region \citep{arav98}. This can pose a problem for accurate determination of the outflow column densities since the uncertainty in the predicted column density becomes large as doublet component ratios approach unity.

A solution to the problem of saturated doublet components is to observe lines with large theoretical optical depth ratios, ensuring that, even at high column densities, the apparent optical depth ratio remains large. An excellent candidate for this kind of study is absorption arising from the n=2 triplet state metastable level in the helium atom. This excited state has an average lifetime of $\sim$2.2 hours \citep{drake71} and is 19.8~eV above the ground state, ensuring that it is predominately populated through recombination of He~{\sc ii}, rather than collisional excitation. As a result, the absorption lines from this level in the optical to near-IR range occurring at 3889~\AA{} and 10\,830~\AA{} are effectively high-ionization transitions. For AGN at higher redshift, a particular advantage of the study of these lines is that they are not contaminated by Ly$\alpha$ forest absorption. This contrasts with the use of P~{\sc v} $\lambda$$\lambda$1118,1128 as a high ionization, high column density-sensitive absorber, whose wavelength is blueward of the Ly$\alpha$ line. The He~{\sc i} singlet level $2^{1}S$ can also be considered metastable, although its lifetime, at $\sim$0.11 seconds, is considerably shorter than that of the $2^{3}S$ level. 

The usefulness of metastable helium in AGN absorption line research was demonstrated by \citet{leighly11}, in which the first example of a He~{\sc i} $\lambda$10\,830 broad absorption line quasar was analysed and from which the outflow properties were constrained. Their investigation used two of the triplet components present in the near-IR at 10\,830~\AA{} and optical at 3889~\AA{}, which have a theoretical optical depth ratio of 23.3:1. A third absorption line representing a transition from the triplet metastable state can be found at 3188~\AA{}. However, this transition is outside of the optical data's spectral range given the low redshift of NGC~4151. Recent studies of AGN metastable helium absorption are also described in \citet{ji15} and \citet{liu15}. The transition from the He~{\sc i} singlet state due to absorption of 20\,587 \AA{} photons can also appear as a weak blueshifted feature. This absorber is not well studied in the literature, however a detection was reported in the near-IR spectrum of NGC~4151 by \citet{iserlohe13} (See \S{}7.2 and fig.~9 in that paper).

This paper investigates the absorption lines resulting from the triplet state absorption components at 3889~\AA{} and 10\,830~\AA{} as well as that due to the transition from the singlet state at 20\,587~\AA{} in an attempt to identify the variability mechanism and constrain the absorber properties. To our knowledge the work described here is the first to examine an AGN outflow using absorption features from all three of the above transitions. This paper is structured as follows: (a) \S{}2 describes the observations taken, (b) \S{}3.1 and \S{}3.2 describe in detail the procedure used to construct the unabsorbed models from which absorption strength is measured, (c) \S{}3.3 details the calculation of column densities, (d) \S{}4 describes the photoionization simulations used, and (e) \S{}5 provides conclusions derived from the observations/simulations. 

\section{Near-IR and optical spectroscopy}

We are using here six epochs of quasi-simultaneous (within 37~days) near-IR and optical spectroscopy for NGC~4151 (see Table~\ref{tab:obsdata}). The near-IR spectroscopy was obtained using the SpeX spectrograph \citep{Ray03} at the NASA Infrared Telescope Facility (IRTF), a 3m telescope on Mauna Kea, Hawai'i, in the short cross-dispersed mode (SXD, $0.8-2.4$ $\mu$m). All data except those from 2010 were obtained through a slit of $0.8\times15''$ giving an average spectral resolution of full width at half maximum (FWHM) $\sim 400$~km~s$^{-1}$. A narrower slit of $0.3\times15''$ was used for the 2010 epoch. The three epochs spanning the years $2004-2007$ were presented in \citet{L11a}, with the $2004$ observation also discussed in \citet{L08a}. The near-IR spectra from 2002 and 2010 were discussed by \citet{Rif06} and \citet{Schnuelle13}, respectively. The recent near-IR spectrum from 2015 was obtained with the refurbished SpeX as part of a variability campaign and will be discussed in detail elsewhere (Landt et al., in prep.). 

The optical spectra, with the exception of the most recent one, were obtained with the FAST spectrograph \citep{Fast98} at the Tillinghast \mbox{1.5 m} telescope on Mt. Hopkins, Arizona, using the 300~l/mm grating and a long-slit of $3''$ aperture. This set-up resulted in a wavelength coverage of $\sim 3720-7515$~\AA~and an average spectral resolution of FWHM $\sim 330$~km~s$^{-1}$. Except for the 2004 data, all spectra were observed at a very low airmass ($\sec~z \sim 1.05$). The May 2004 spectrum was observed at an airmass of $\sec~z \sim 1.3$ and so the flux loss due to atmospheric differential refraction is expected to be $\sim 20\%$ at the observed wavelength of He~{\sc i}~$\lambda$3888 relative to that at wavelengths $\ge5000$~\AA~\citep{Fil82}. All FAST optical spectra were retrieved from the FAST archive. The optical spectrum from 2015 was obtained with the ISIS dual-arm spectrograph at the William Herschel \mbox{4.2 m} telescope on La Palma, Canary Islands, using the 600~l/mm and 316~l/mm gratings for the blue and red arm, respectively, and a $1''$ aperture long-slit. This set-up resulted in a total wavelength coverage of $\sim 3690-8800$~\AA~and an average spectral resolution of FWHM $\sim 75$~km~s$^{-1}$ and $\sim 145$~km~s$^{-1}$ in the blue and red arm, respectively. The spectrograph slit was rotated to align with the parallactic angle, in order to minimise the effect of atmospheric refraction on the analysis. Reduction of the optical data was performed within {\sc iraf}, while the IRTF reduction was carried out using \emph{Spextool} \citep{cushing04}. Before any analysis of the data was performed, all spectra were adjusted to the object rest-frame.  

\begin{table*}
\begin{center}
\caption{Quasi-simultaneous near-IR and optical observations.}
\begin{tabular}{l l l l l l}
\hline
\textbf{Optical Observations}& & & & & \\
Epoch&Date&Position angle&Airmass&Exp. time&Signal/Noise at 4000~\AA{}\\
 & &$^{\circ}$& &(seconds)& \\
\hline
1&2002 Apr 11&70&1.04&3 $\times$ 30&19.1\\
2&2004 May 28&90&1.30&2 $\times$ 30&10.1\\
3&2006 May 29&90&1.03&2 $\times$ 30&10.1\\
4&2007 Feb 9&90&1.05&2 $\times$ 30&12.0\\
5&2010 Feb 18&90&1.05&3 $\times$ 60&26.9\\
6$^{\dagger}$&2015 Mar 14&254&1.10&3 $\times$ 180&16.2\\
\hline
\textbf{Near-IR Observations}& & & & & \\
Epoch&Date&Position angle&Airmass&Exp. time&Signal/Noise at 11\,000~\AA{}\\
 & &$^{\circ}$& &(seconds)& \\
\hline
1&2002 Apr 23&130&1.10&1800&243\\
2&2004 May 23&90&1.13&6 $\times$ 120&28.0\\
3&2006 Jun 12&90&1.22&10 $\times$ 120&78.4\\
4&2007 Jan 24&90&1.45&8 $\times$ 120&29.0\\
5&2010 Feb 27&188&1.06&10 $\times$ 180&240\\
6&2015 Feb 5&223&1.08&8 $\times$ 120&28.3\\
\hline
\label{tab:obsdata}
\end{tabular}
\end{center}
$^{\dagger}$ This observation was undertaken with the WHT, all other optical spectra were obtained from the Tillinghast telescope.\\
\end{table*}

\section{Analysis}

\subsection{Spectral adjustments}

The six spectra were observed at optical and near-IR wavelengths in order to span the $2^{3}S$ He~{\sc i} emission and absorption features resulting from two of the components, at 3889~\AA{} and 10\,830~\AA{}. In both wavebands small additive wavelength shifts were made to account for differences in wavelength calibration across the various epochs by aligning the spectra to the wavelengths of strong skylines prior to correction for redshift. Before any analysis of the spectra were undertaken, small variations in the effective resolution were removed by identifying the spectrum across both sets of observations with the broadest narrow features. All other spectra were then convolved with a Gaussian of unit area and appropriate FWHM to match the resolution of this spectrum ($\sim$400~km~s$^{-1}$) while preserving the area of emission and absorption features. 

\subsection{The unabsorbed spectrum}

\begin{figure}
\centering
\resizebox{\hsize}{!}{\includegraphics[angle=0,width=8cm]{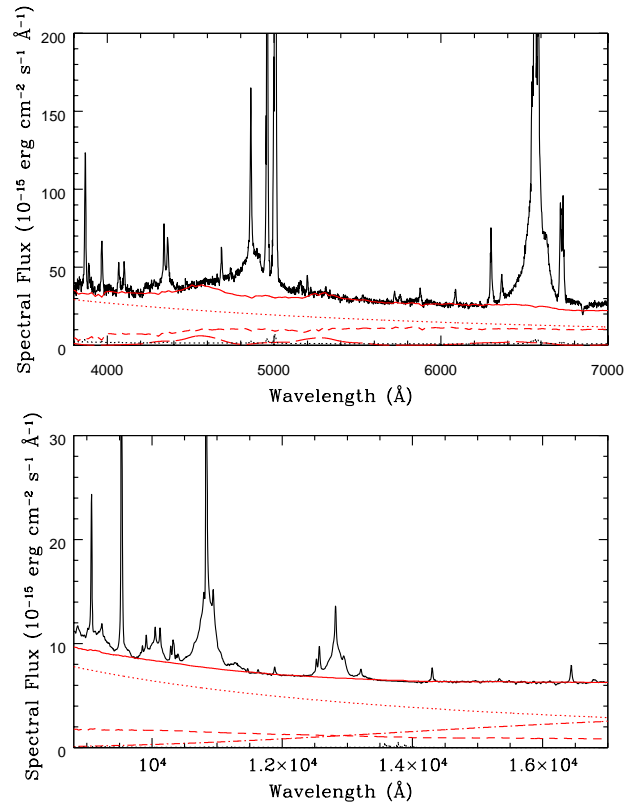}}
\caption{Continuous emission fits to the optical spectrum (top panel) and the near-IR spectrum (bottom panel) at epoch~1. The AGN continuum (red dotted line), host galaxy (red short-dashed line), Fe~{\sc ii} broad emission (red long-dashed line) and blackbody torus emission (red dot-dashed line) are indicated along with the observed spectrum and its associated error spectrum (black solid and black dotted lines respectively). The total continuous emission model is indicated by the red solid line. The near-IR errors are barely visible due to the high S/N.}
\label{fig:confit}
\end{figure}

Reconstruction of the unabsorbed spectrum is required in order to measure the depth of the absorption features. This process is complicated by the overlap of emission from several different sources, namely the AGN power-law continuum, the broad line region (BLR) including a continuum of blended broad Fe~{\sc ii} emission, the NLR, the torus and the host galaxy. To deal with the relatively featureless and continuous aspects of this emission, the aforementioned Fe~{\sc ii} continuum together with the host galaxy starlight and AGN power-law contributions were fitted simultaneously to the same relatively line-free regions of the spectra at each epoch using the {\sc iraf} \emph{specfit} package, which was also used in all subsequent spectral fitting described in this paper. For the near-IR spectra, an additional contribution from the dusty torus was modelled using a blackbody. The Fe~{\sc ii} model was scaled from the template obtained from I Zw 1 by \citet{veroncetty04}, which was also Gaussian-blurred to approximate the width of the broad emission lines in the spectrum. The host galaxy contribution was built from the Sa and Sb-type spiral galaxy templates listed in the SWIRE library \citep{polletta07}. Epochs 2, 3 and 4 have estimated galaxy fluxes, derived from HST observations, listed in \citet{L11a}. Galaxy templates were therefore normalised to match these fluxes prior to running the \emph{specfit} fitting routine, minimising the problem of degenerate fitting. Epochs without such observations used an average value, corrected when necessary for differences in slit width. An example of the resulting power-law, galaxy starlight, dust torus and Fe~{\sc ii} emission models fitted to the observed spectra is illustrated in Fig.~\ref{fig:confit}.

\subsubsection{Reconstruction of the He~{\sc i} 10\,830\AA{} line}

\begin{figure}
\centering
\resizebox{\hsize}{!}{\includegraphics[angle=0,width=8cm]{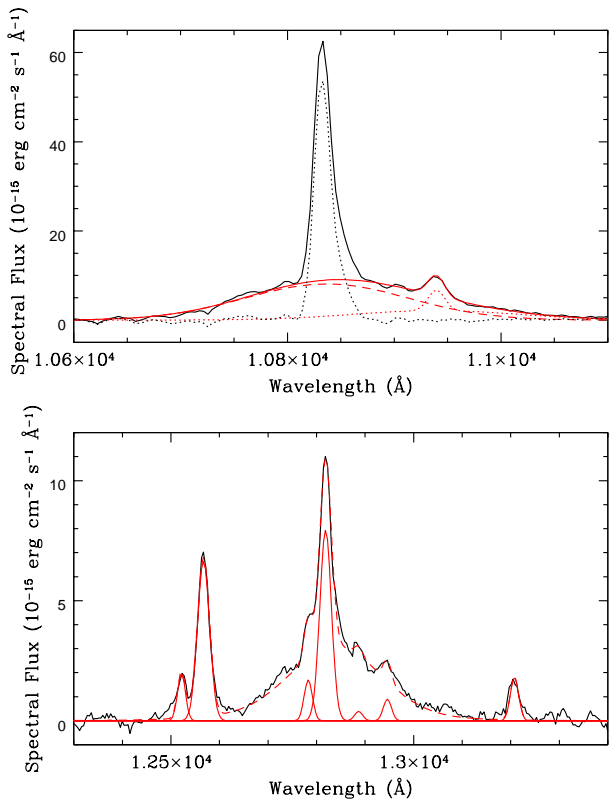}}
\caption{\emph{Top panel:} Isolation of He~{\sc i} narrow component (black dotted line) from epoch~6 by subtraction of the He~{\sc i} broad component (red dashed line) and Pa$\gamma$ broad+narrow line (red dotted line). The total subtracted model is represented by the red solid line, while the black solid line is the observed He~{\sc i} blend. \emph{Bottom panel:} Epoch~6 observed Pa$\beta$ blend (solid black line) and Gaussian-modelled narrow lines (solid red lines). The red dashed line indicates the total line-blend model. Modelled narrow lines were subtracted to leave a broad template at epochs~1--5. Both panels show spectra after subtraction of the underlying continuous emission.}
\label{fig:nirhe1rec}
\end{figure}

The near-IR metastable $2^{3}S$ helium line is reconstructed by isolating unabsorbed narrow and broad components and fitting these to the He~{\sc i} $\lambda$10\,830 complex. The process is complicated by  blending with the nearby Pa$\gamma$ line. It was noted that the epoch~6 near-IR spectrum showed no evidence of absorption features, therefore the narrow component was isolated and used as a template for all other epochs. The broad lines in the epoch~6 spectrum are relatively weak and symmetrical, therefore it was a simple process to isolate this narrow component by fitting the broad components of both the He~{\sc i} $\lambda$10\,830 line and the Pa$\gamma$ line with two Gaussians each. At epochs 1--5, the broad emission components of He~{\sc i} and Pa$\gamma$ are not symmetrical. To deal with this, a broad emission template was extracted from the Pa$\beta$ complex at each of these epochs by first identifying blended narrow components in the simpler epoch~6 spectrum using appropriate Gaussian models (listed in Table~\ref{tab:pabblends}) and then subtracting them out at other epochs after appropriate scaling. Identification of the narrow component of He~{\sc i} $\lambda$10\,830 and the Pa$\beta$ blends is illustrated in Fig.~\ref{fig:nirhe1rec}.

\begin{table*}
\begin{center}
\caption{Narrow lines subtracted from Pa$\beta$ blend.}
\begin{tabular}{l l}
\hline Blended narrow line&Central wavelength\\
 &(\AA{})\\
\hline
$[$S~{\sc ix}$]$&12\,520\\
$[$Fe~{\sc ii}$]$&12\,567\\
$[$Fe~{\sc ii}$]$&12\,790\\
Pa$\beta$&12\,818\\
Unknown, possibly Fe~{\sc ii} or Ne~{\sc i}&12\,888\\
$[$Fe~{\sc ii}$]$&12\,943\\
$[$Fe~{\sc ii}$]$&13\,205\\
\hline
\end{tabular}
\label{tab:pabblends}
\end{center}
\end{table*}

\begin{figure}
\centering
\resizebox{\hsize}{!}{\includegraphics[angle=0,width=8cm]{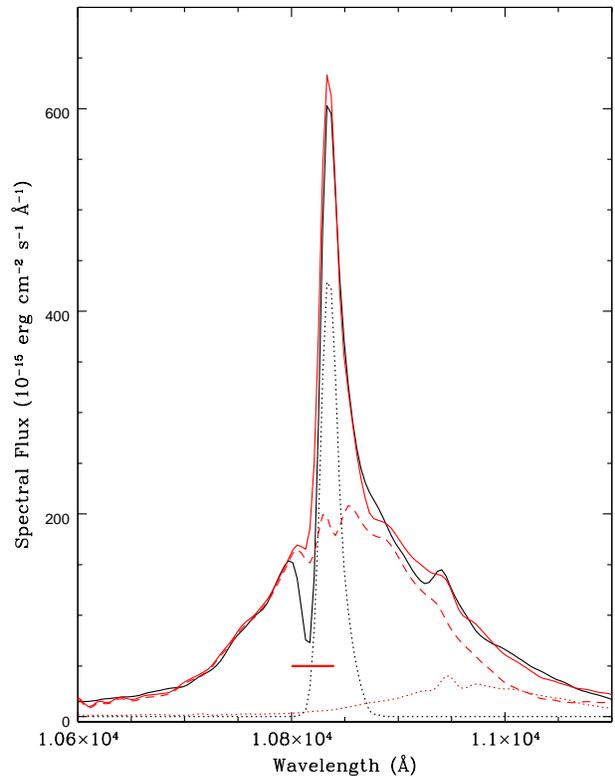}}
\caption{Fitted components to the He~{\sc i} $\lambda$10\,830 complex at epoch~5 after subtraction of the continuous emission. The He~{\sc i} narrow component is represented by the black dotted line, the He~{\sc i} broad component by the red dashed line, and the Pa$\gamma$ broad+narrow emission by the red dotted line. The total unabsorbed model is represented by the red solid line, while the black solid line is the observed He~{\sc i} blend. Both broad components were derived from the Pa$\beta$ broad template. The horizontal red line indicates the wavelengths spanned by an absorption feature.}
\label{fig:ep5blend}
\end{figure}

These templates, together with a narrow Gaussian representing the narrow component of Pa$\gamma$, were simultaneously scaled to the observed blend to produce a reconstruction of the unabsorbed He~{\sc i} $\lambda$10\,830 complex. An example of the individual components used to fit the blend is illustrated in Fig~\ref{fig:ep5blend}.

\subsubsection{Reconstruction of the He~{\sc i} 20\,587~\AA{} line}

\begin{figure}
\centering
\resizebox{\hsize}{!}{\includegraphics[angle=0,width=8cm]{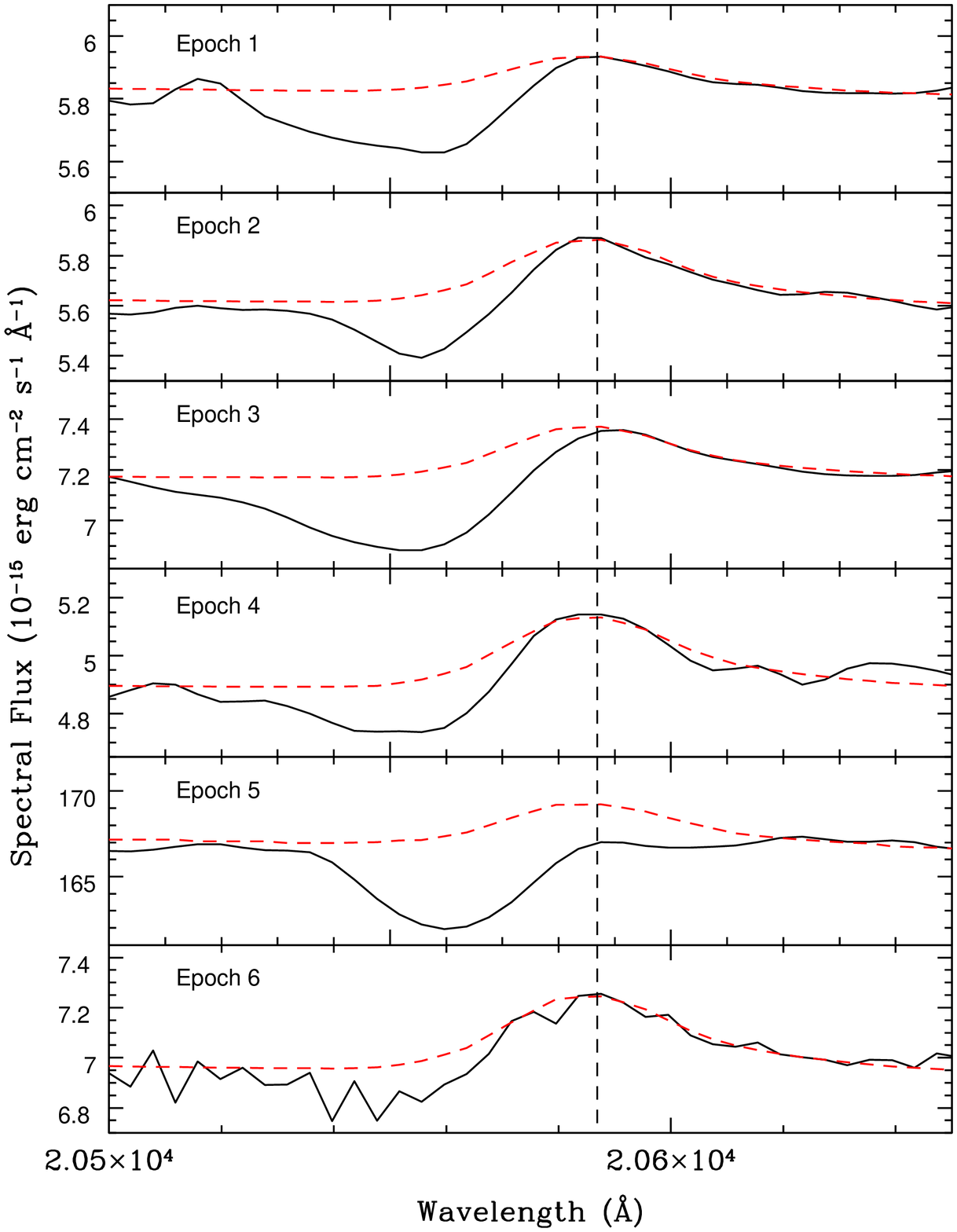}}
\caption{Observed spectrum (solid black line) and unabsorbed reconstruction (dashed red line) in the spectral region near He~{\sc i} $\lambda$20\,587 (line centre indicated by the vertical black dashed line).}
\label{fig:he20587unabs}
\end{figure}

The helium emission line at 20\,587~\AA{} is very weak compared to the triplet state emission line in the near-IR spectrum, as is the corresponding blueshifted absorption. To reconstruct this emission line at each epoch, the narrow component of the He~{\sc i} $\lambda$10\,830 line was interpolated onto the wavelength of the 20\,587~\AA{} line and underwent scaling and linear flux shift (at most 1 wavelength bin) to match the red (unabsorbed) side of the emission line by means of chisquare minimisation. No broad component of this line was discernible, therefore no attempt was made to produce a broad component as part of the reconstruction. An exception to this method was that used for the epoch~5 reconstruction, where the absorption is sufficiently strong that the narrow emission line is lost in the absorption trough. In this case, the scale factor between the narrow component at 10\,830~\AA{} and 20\,587~\AA{} was assumed to be the same as that at epoch~6, since changing line ratios in response to ionizing continuum changes in the relatively extended narrow line region only happen over relatively long time periods. The final reconstructed line+continuum emission templates at each epoch are indicated in Fig.~\ref{fig:he20587unabs}. It proved difficult to fit an accurate continuum template to this region of the spectrum, so the continuum was assumed to be a flat flux level at the line base. The resultant lack of deconvolution between the various components (power-law, torus and host galaxy emission), together with the absorber's singlet structure, means only lower column density limits can be calculated using this line.

\subsubsection{Optical spectral region $<$ 4000 \AA{}}

The unabsorbed emission at the blue end of the optical spectra is a complex blend of several different contributing sources. It is necessary to model the unabsorbed emission in this region accurately due to the weakness of the He~{\sc i} $\lambda$3889 feature. Hydrogen emission includes higher order Balmer series lines as well as the Balmer continuum discontinuity at 3646~\AA{} which, given its observed wavelength shift and high degree of broadening, also contributes excess flux to this region. The narrow components of the Balmer series lines are straightforward to calculate, as it can be assumed that they form in a low-density environment where the dominant process generating the lines is Case B recombination. 

\begin{figure}
\centering
\resizebox{\hsize}{!}{\includegraphics[angle=0,width=8cm]{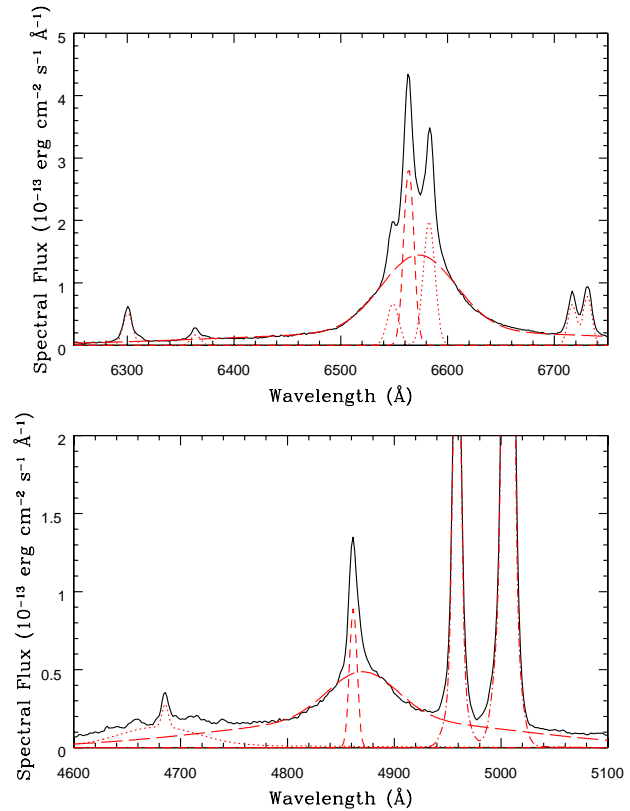}}
\caption{Gaussian fits to the spectral region of H$\alpha$ (top panel) and H$\beta$ (bottom panel) at epoch~5. In both panels the solid black line represents the observed spectrum with the underlying continuous emission subtracted. The short-dashed and long-dashed red lines represent the Balmer line narrow and broad components respectively. The dotted red lines in the top panel represent the blended narrow emission lines, while in the bottom panel the dotted red line represents the blended He~{\sc ii} emission feature. The dot-dashed red lines in the bottom panel represent the blended [O~{\sc iii}] doublet components.}
\label{fig:halphbet}
\end{figure}

The narrow line intensities can be calculated by referring to the theoretical H$\beta$ ratios from \citet{osterbrock06} and assuming the widths are the same as for the narrow component of H$\beta$. The H$\alpha$ and H$\beta$ emission profiles were modelled as three Gaussians (two for the broad component and one for the narrow component), an example fit is illustrated in Fig.~\ref{fig:halphbet}. It was found that the narrow component H$\alpha$/H$\beta$ flux ratios were always greater than those predicted by theory. This was assumed to be due to intrinsic dust extinction of the form observed in the SMC bar with $R_{V}$=2.74 \citep{gordon03}. To correct for this reddening, test $E(B-V)$ values were incrementally varied and applied to the theoretical ratios until the observed H$\alpha$/H$\beta$ ratios were correctly predicted. The epoch~1 optical spectrum was excluded from the subsequent absorption line analysis due to the multi-peaked nature of the H$\alpha$ line, rendering an accurate fit very difficult to attain. This was not expected to alter the conclusions of this investigation as no obvious absorption is present at this epoch in either spectrum.

The broad components of these lines do not from in a low-density environment, meaning that electron impact excitation must be taken into account when calculating their intensities. In \citet{popovic03}, applying Boltzmann plots to broad Balmer line intensities revealed that the broad emission line regions of several AGN, including NGC~4151, were in partial local thermodynamic equilibrium (PLTE). We make the assumption of PLTE in this case and use Equation~2 in \citet{popovic03} to calculate the broad component intensities based on the strengths of the lower order lines, assuming that the widths are the same as that of the broad H$\beta$ component. The Balmer continuum can be modelled by relating the Balmer continuum jump at 3646~\AA{} to the electron temperature in the line forming region, assuming the plasma is optically thin and has a single electron temperature \citep{grandi82,wills85}. This results in the following equation 

\begin{equation}
\label{eqn:balmedge}
F_{\nu}^{\rm BC}=F_{\nu}^{\rm BE}e^{\left(-h-\nu{}_{\rm BE}\right)/\left(kT_{e}\right)}\, ,
\end{equation}

\noindent where $F_{\nu}^{BC}$ is the Balmer continuum flux at frequency $\nu$, $F_{\nu}^{BE}$ is the flux at the Balmer edge, $k$ is the Boltzmann constant, $h$ is Planck's constant and $T_{e}$ is the electron temperature. The study of \citet{jin12} indicated that, in order to get an acceptable Balmer continuum fit to the spectrum of PG~1427+480, the position of the discontinuity needed to be shifted redward by 100~\AA{} and broadened by 6000~km~s$^{-1}$. A visual inspection of the Balmer edge calculated at each epoch with no broadening or shifting indicated a similar process was required for our optical spectra. A possible explanation for this wavelength shift could be Stark broadening if the electron density in the Balmer continuum emission region is much higher than in the BLR \citep{pigarov98}, suggesting this region may lie in the illuminated accretion disc rather than the BLR clouds \citep{jin12,collinsouffrin90}. The Balmer continuum was fitted to our spectra by simultaneously and iteratively shifting and broadening the Balmer jump until a good fit was achieved. A shift of $\sim$100~\AA{} and a broadening of $\sim$24\,000~km~s$^{-1}$ were found to give the best fit for the Balmer continuum.

\begin{figure}
\centering
\resizebox{\hsize}{!}{\includegraphics[angle=0,width=8cm]{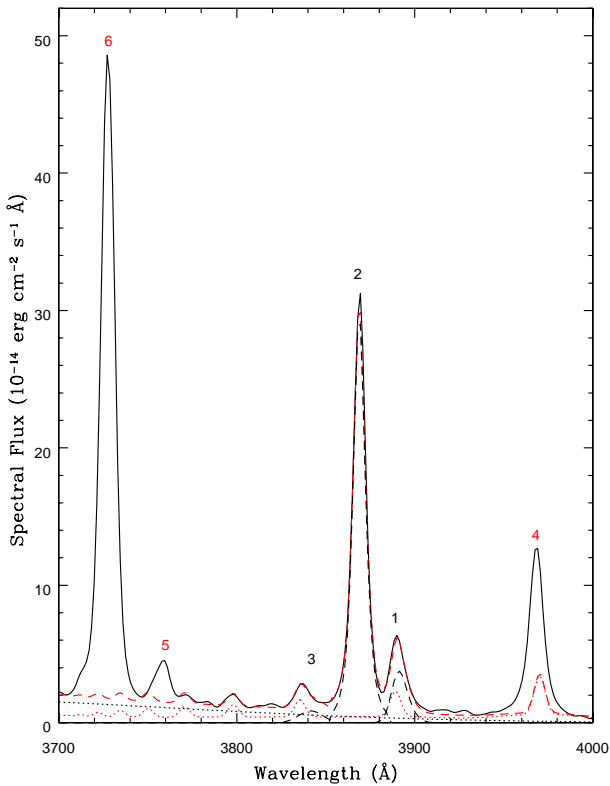}}
\caption{Spectral fitting in the $<$4000~\AA{} region at epoch~6. The solid black line is the observed spectrum with the continuous emission components (power-law, galaxy starlight and broad Fe~{\sc ii}) subtracted. The dotted black line is the broadened and shifted Balmer continuum emission model and the dotted red line is the Balmer BLR+NLR emission model. The dashed black lines underneath labels '1' and '2' are the reconstructed He~{\sc i}~$\lambda$3889 and [Ne~{\sc iii}]~$\lambda$3868 emission lines respectively, while that labelled '3' is an unknown line modelled with a broad Gaussian component. The red dashed line is the total reconstructed spectrum made up of the sum of all aforementioned emission models. Labels '4', '5' and '6' indicate those lines sufficiently far from the helium line that reconstruction was not deemed necessary.}
\label{fig:bcregion}
\end{figure}

\begin{figure}
\centering
\resizebox{\hsize}{!}{\includegraphics[angle=0,width=8cm]{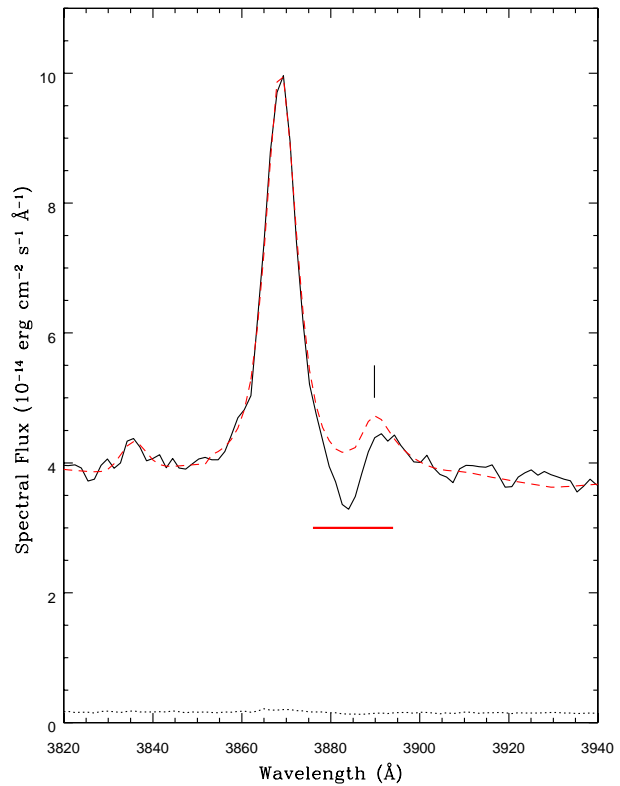}}
\caption{Epoch~5 total optical spectrum (solid black line) and its total reconstruction (red dashed line) near the He~{\sc i} $\lambda$3889 line. The horizontal solid red line indicates the wavelength span of the absorber, while the vertical solid black line indicates the location of the He~{\sc i} emission line centroid. The dotted black line indicates the errors on the observed spectrum.}
\label{fig:ep5rec}
\end{figure}

Once an accurate model of the hydrogen emission has been achieved, this must be added to the He~{\sc i} $\lambda$3889 emission complex to create a complete ``pseudocontinuum'' from which the depth of the absorption feature can be measured. This feature is blended with the [Ne~{\sc iii}] $\lambda$3868 line, which therefore also needs to be reconstructed. Since, as already discussed, the epoch~6 near-IR spectrum shows no evidence of metastable He~{\sc i} absorption, the narrow component of the 10\,830 \AA{} line can be used as a template for the He~{\sc i}~$\lambda$3889 line. No broad component for this line is required as it is so weak as to be indiscernible. Given the large ratio between the oscillator strengths of the near-IR and optical lines and the lack of absorption in the near-IR line, it is unlikely that epoch~6 shows any metastable helium absorption corresponding to the optical transition. Therefore, the epoch~6 near-IR He~{\sc i} narrow emission line was scaled to match the red (unblended) side of the corresponding optical emission profile. The [Ne~{\sc iii}]+He~{\sc i} blend could then be fitted at epoch~6 by adding to this He~{\sc i} template a [Ne~{\sc iii}] template, generated by inverting the unblended blue side of the observed line to reconstruct the blended red side. These two templates were then added to the calculated Balmer emission components along with a broad ($\sim$3000~km~s$^{-1}$) Gaussian to account for unknown emission at 3840 \AA{}. This completes the total unabsorbed reconstruction at epoch~6, as shown in Fig.~\ref{fig:bcregion}. Epochs 1--5 were fitted using the calculated Balmer emission together with appropriate scaling of the epoch~6 [Ne~{\sc iii}] and He~{\sc i} templates. The total unabsorbed reconstruction of the epoch~5 spectrum (which contains an absorption feature) is shown in Fig~\ref{fig:ep5rec}.

\subsection{Singlet and triplet state metastable helium column densities}

\begin{figure}
\centering
\resizebox{\hsize}{!}{\includegraphics[angle=0,width=8cm]{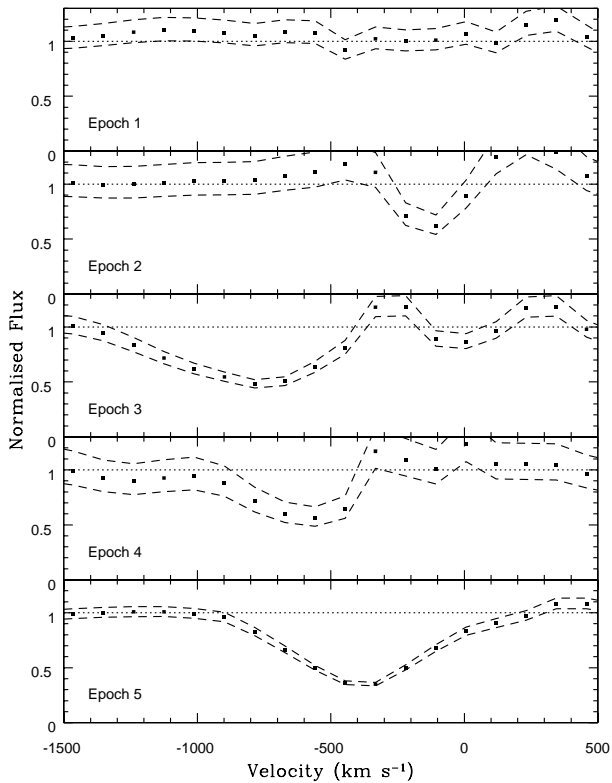}}
\caption{He~{\sc i} $\lambda$10\,830 absorber profiles across epochs 1 to 5. Black squares indicate the residual flux at each velocity bin. Dashed lines show the envelope of values allowed by the fractional error in the emission profiles.}
\label{fig:irnorms}
\end{figure}

The simplest method of measuring outflow column densities is to assume that the absorber is unsaturated and covers 100 per cent of the emission region. This requires a simple integration across the absorber profile \citep{savage91} given by

\begin{equation}
\label{eqn:dicalc}
N_{ion}=\frac{m_{e}c}{\pi{}e^{2}f\lambda{}}\int \tau{}\left(v\right)dv\, ,
\end{equation}

\noindent where $N_{ion}$ is the ionic column density, $v$ is the velocity relative to the laboratory rest-frame wavelength of the transition, $m_{e}$ is the electron mass, $c$ is the speed of light, $e$ is the elementary charge, $f$ is the oscillator strength, $\lambda$ is the laboratory rest-frame wavelength and $\tau(v)$ is the optical depth at a given velocity. For this investigation wavelength measurements were converted to velocities relative to emission line centre. Sources of error include the point-to-point uncertainty in the flux values and a fractional error associated with the reconstructed spectrum. This fractional error was derived from the root-mean-square of the normalised difference values between the reconstructed and observed spectra at $\sim$1300~km~s$^{-1}$ either side of the spectral region spanned by velocity bins which show absorption in at least one epoch. Calculations were performed separately for the optical and near-IR spectra at epochs 2--6 and 1--6 respectively. The emission region is assumed to be that generating the modelled power-law continuum emission over the span of the absorption feature (see \S{}3.3.1 for justification). The near-IR emission-region-normalised $2^{3}S$ profiles at epochs 1 to 5 are shown in Fig.~\ref{fig:irnorms} (epoch~6 has zero absorption by definition). Absorption is apparent at epochs 3--5, with the epoch~5 absorber occupying, on average, a region of lower (less negative) outflow velocity compared to epochs 3 and 4. This suggests a possible separate physical origin of the epoch~5 absorber compared to earlier epochs, especially at $v$$>$$-$400~km~s$^{-1}$.

\begin{figure}
\centering
\resizebox{\hsize}{!}{\includegraphics[angle=0,width=8cm]{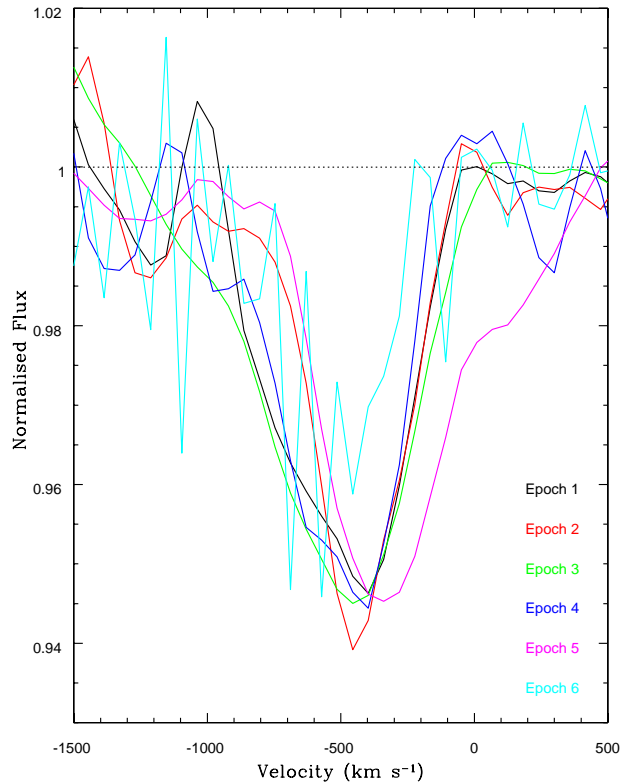}}
\caption{He~{\sc i} $\lambda$20\,587 absorber profiles across epochs 1 to 6. Each profile is colour-coded by epoch as shown in the bottom right.}
\label{fig:he20587n}
\end{figure}

Using the unabsorbed reconstructions in the spectral region of the He~{\sc i} $\lambda$20\,587 line shown in Fig.~\ref{fig:he20587unabs}, a value of the column density in this state can be estimated from the normalised spectral profile of the absorbed region (excluding the narrow emission line as the NLR is probably uncovered by the absorber). Since the He~{\sc i} transition at 20\,587 \AA{} is a singlet, only a lower limit for the column density can be obtained using direct integration of the profile, which will be an underestimate if partial coverage and saturation are present. However, we would expect these limits to be close to the true value given the shallowness of the absorber \citep{leighly11}. As mentioned in \S{}3.2.2, column densities also need to be lower limits due to a lack of separation of the various emission sources at the wavelengths spanned by the singlet state absorber. The normalised absorption profiles are shown in Fig.~\ref{fig:he20587n}. The derived column densities for both $2^{3}S$ (from the 3889 \AA{} line and the 10\,830 \AA{} line) and $2^{1}S$ state helium are recorded in Table~\ref{tab:dintcols}.

\begin{table*}
\begin{center}
\caption{Apparent column densities calculated at each epoch over the optical and near-IR absorption regions. Epoch~1 is excluded from the optical analysis due to the lack of a reconstructed profile. Zero values include negative values within the error limits.}
\begin{tabular}{l l l l}
\hline Epoch&Optical $2^{3}S$ column density&Near-IR $2^{3}S$ column density&$2^{1}S$ column density (lower limit)\\
 &(10$^{13}$ cm$^{-2}$)&(10$^{13}$ cm$^{-2}$)&(10$^{13}$ cm$^{-2}$)\\
\hline
1&N/A&0$\pm$0.98&0.148$\pm$0.035\\
2&77.7$\pm$121&0.12$\pm$1.47&0.134$\pm$0.046\\
3&88.1$\pm$120&2.73$\pm$0.52&0.177$\pm$0.028\\
4&79.9$\pm$40.3&1.45$\pm$0.59&0.146$\pm$0.068\\
5&28.3$\pm$5.87&3.44$\pm$0.30&0.176$\pm$0.021\\
6&0$\pm$108&0$\pm$1.17&0.111$\pm$0.041\\
\hline
\end{tabular}
\label{tab:dintcols}
\end{center}
\end{table*}

The unreliability of the optical measurements (due to the quality of the spectrum, the weakness of the continuum and the weakness of any 3889~\AA{} He~{\sc i} absorber) is clear from these values for all but epoch 5, which is the only observation having a column density of greater than 2$\sigma$ significance (4.8$\sigma$). The strength of the near-IR absorber allows more reliable measurements to be made of absorber variability, which will be discussed further in \S{}5. The $2^{1}S$ column density values are mutually consistent across the epochs, so no variability in this absorber is identifiable. This absorber is weak and the position of underlying emission is uncertain, so the lack of variability is not considered indicative.

\subsubsection{Partial covering analysis of epoch~5}

The significance of the optical absorber at epoch~5 allows a partial coverage analysis to be performed at this epoch by using both the optical and near-IR absorbers. We use a method similar to that adopted by \citet{leighly11}, applying a pure partial coverage model (PPC) to the optical and near-IR $2^{3}S$ He~{\sc i} absorption components. Although the investigation of that paper also included a power-law inhomogeneous model, this yielded column densities not significantly different from the PPC model in that study, and hence it is not used in this investigation. The PPC model was originally applied to quasar absorption lines \citep{hamann97,arav99} and relates the observed flux density within an individual wavelength bin to the total flux from the emission source and the covering fraction as follows:

\begin{equation}
\label{eqn:ppc1}
\frac{I_{{\rm app}}}{I_{{\rm 0}}}=Ce^{-\tau}+\left(1-C\right)\, ,
\end{equation}

\noindent where $I_{{\rm app}}$ is the apparent flux, $I_{{\rm 0}}$ is the flux from the emission region, $C$ is the fraction of the total emission covered by the absorber and $\tau$ is the optical depth. In order to compare the normalised depth of each component, the normalised near-IR absorption line spectrum was interpolated into the velocity space of the normalised optical spectrum with velocity bin width $\sim$113~km~s$^{-1}$. By taking into account the 23.3 times greater optical depth of the near-IR component as compared to the optical component, the strengths can be compared using the equation

\begin{equation}
\label{eqn:ppc2}
\left(\frac{I_{{\rm op}}-1}{C}+1\right)^{23.3}-\frac{\left(I_{{\rm IR}}-1\right)}{C}-1=0\, ,
\end{equation}

\noindent where $I_{{\rm op}}$ and $I_{{\rm IR}}$ are the normalised fluxes of the optical and near-IR components respectively. 

Before performing the calculations it is necessary to identify the emission region which is covered by the absorber along the line-of-sight (providing $I_{{\rm 0}}$). An outflow associated with the active nucleus will not absorb a significant portion of the galaxy flux. The optical absorber is deeper than the combined BLR and torus (blackbody) emission, so sole coverage of one or both of these regions is ruled out. This leaves three possibilities for the absorbed region: power-law continuum only, continuum+BLR and continuum+BLR+torus. Identification of the best option is achieved using the condition $I_{{\rm op}}$$\geq$$I_{{\rm IR}}$$\geq$$I_{{\rm op}}$$^{23.3}$, which is necessary for the validity of Equation~\ref{eqn:ppc2}. Assuming continuum coverage only, all nine velocity bins showing significant absorption in the stronger (near-IR) component are valid, while the other possibilities (continuum+BLR and continuum+BLR+torus) result in three and five bins violating the condition respectively. The emission region is therefore assumed to be that producing the power-law continuum, which is used as $I_{{\rm 0}}$ in all calculations in this paper involving $2^{3}S$ helium.

\begin{figure}
\centering
\resizebox{\hsize}{!}{\includegraphics[angle=0,width=8cm]{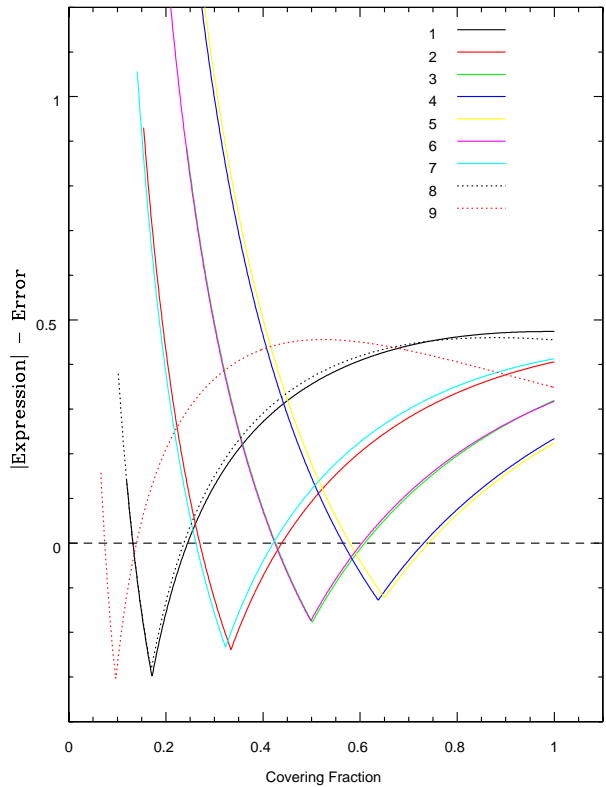}}
\caption{Curves representing the magnitude of the left-hand-side of Equation~\ref{eqn:ppc2}, minus the associated error, for covering fractions between 0 and 1 at each valid velocity bin (line number labels in top right correspond to those in Fig.~\ref{fig:ep5profile}). Where each curve crosses zero (two points) are taken as the lower and upper limits of $C$ at that bin.}
\label{fig:cfracs}
\end{figure}

\begin{figure}
\centering
\resizebox{\hsize}{!}{\includegraphics[angle=0,width=8cm]{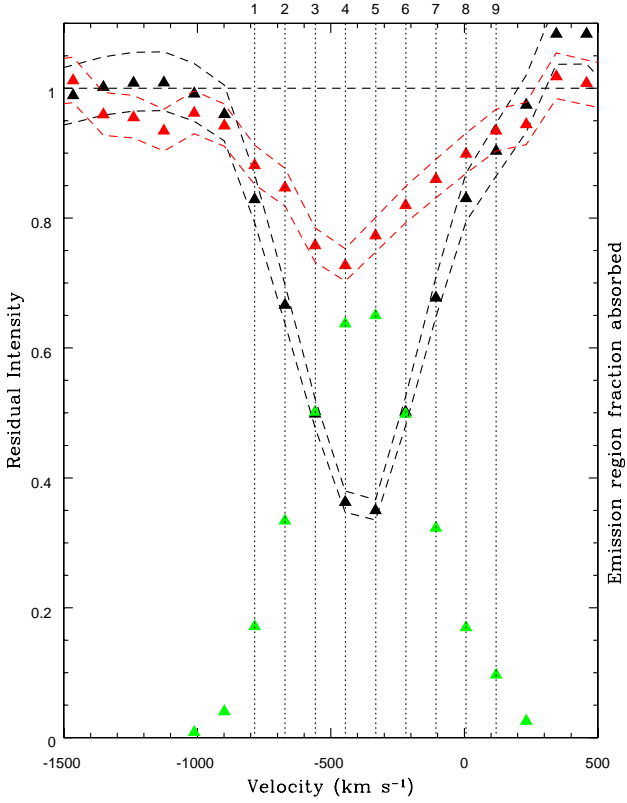}}
\caption{Normalised absorption profiles of triplet state metastable helium absorption at 3889~\AA{} (red triangles) and 10\,830~\AA{} (black triangles), representing the residual intensity at each velocity bin. Velocity bins which satisfy $I_{{\rm op}}$$\geq$$I_{{\rm IR}}$$\geq$$I_{{\rm op}}$$^{23.3}$ are located at velocities marked by dotted vertical lines and are assigned numerical labels in order of increasing (positive direction) velocity, with negative velocities indicating outflow from the emission source. Green triangles represent the continuum emission absorbed at each velocity bin as derived from the near-IR spectrum. Dashed lines show the envelope of values allowed by the fractional error in the emission model.}
\label{fig:ep5profile}
\end{figure}

The calculation of column density relies on finding a solution for $C$ in Equation~\ref{eqn:ppc2}. Unlike the case of a UV absorber such as Si~{\sc iv} (with doublet optical depth ratio 2:1), Equation~\ref{eqn:ppc2} has no analytic solution. Therefore, a numerical test is performed by varying $C$ in steps of 0.001 between 0 and 1 at each velocity bin. At each step, the error associated with the expression on the left-hand-side of Equation~\ref{eqn:ppc2} is calculated. Solutions of $C$ within error limits therefore correspond to those steps where the magnitude of the expression minus its error is less than zero, as illustrated in Fig.~\ref{fig:cfracs}. This test is carried out at all nine velocity bins which satisfy $I_{{\rm op}}$$\geq$$I_{{\rm IR}}$$\geq$$I_{{\rm op}}$$^{23.3}$, with the normalised absorption profiles shown in Fig.~\ref{fig:ep5profile}. It is clear, given the large theoretical component ratio, that the near-IR absorber must be saturated across the region spanned by the nine velocity bins. In such a case the absorber profile is determined by the covering fraction. The column density can be calculated in a similar way to Equation~\ref{eqn:dicalc} as follows

\begin{equation}
\label{eqn:ppc3}
N_{{\rm ion}}=\frac{m_{{\rm e}}c}{\pi{}e^{2}f\lambda{}}\int \tau\left(v\right)C\left(v\right)dv\, ,
\end{equation}

\noindent giving a triplet state metastable helium column density of 3.47$^{+0.99}_{-1.05}$$\times$10$^{14}$~cm$^{-2}$. This is therefore a well constrained absorber suitable for further analysis (see \S{}4.)

\section{Photoionization simulations}

\subsection{{\sc cloudy} simulations}

In order to locate the epoch~5 triplet state absorber in ionization parameter, number density and column density space, grids of photoionization simulations were calculated. The photoionization simulations calculated for this paper were performed using {\sc cloudy} version c13.02 \citep{ferland13}, for which the setup was similar to that described in paragraph 1 of \S{}3.1 in \citet{leighly11}. Simulations were available which utilised a triplet state He~{\sc i} column density of $n_{{\rm He}}(2^{3}S)$=3.67$\times$10$^{14}$~cm$^{-2}$, well within the errors on the epoch~5 absorber determined from the PPC analysis. These simulations are those discussed in this section.

The ionization parameter ($U$) is a useful quantity for measuring the ratio of Lyman continuum photons to hydrogen atoms as follows

\begin{equation}
\label{eqn:ionparam}
U=\frac{Q(H)}{4\pi{}r^{2}n_{{\rm H}}c}\, ,
\end{equation}

\noindent where $Q(H)$ is the number of Lyman continuum photons produced by the ionizing source per second, $r$ is the radial distance from the photon source and $n_{{\rm H}}$ is the hydrogen number density. We use an input spectral energy distribution (SED) that follows the shape of the AGN continuum described in \citet{korista97b} and we adopt solar abundances from \citet{grevesse98}, \citet{holweger01}, and \citet{allendeprieto01,allendeprieto02}. This continuum is assigned to the {\sc cloudy} calculations by using the ``AGN Kirk'' command.

Balmer absorption was detected in NGC 4151 by \citet{hutchings02}, arising from the H$\beta$ transition. In \citet{leighly11}, the lack of Balmer absorption in the quasar FBQS J1151+3822 was used to set an upper limit on the n=2 hydrogen state column density. The spectral region of H$\beta$ at epoch~5 was examined to determine whether or not H$\beta$ absorption was present. This employed the multiple Gaussian reconstruction used previously (see Fig.~\ref{fig:halphbet}) as the unabsorbed reconstruction. No absorber is apparent from a visual inspection of the spectrum and the reconstruction. This lack of absorption can be used to set limits on the allowed ionization parameter and total hydrogen column density of the absorber as described in \S{}4.3.

Our calculations require two runs of the code over each grid. The first run is used to find the thickness of the absorbing cloud to be integrated through to reach the spectroscopically-determined column density of helium in the $2^{3}S$ metastable state at epoch~5. This allows the total hydrogen ($N_{{\rm H}}$) and $2^{1}S$ state helium column densities to be calculated. The second run integrates through to the $N_{{\rm H}}$ values determined from the first run to find the hydrogen level populations, which can then be used to determine if a particular gridpoint exceeds the allowed Balmer absorption. Spectroscopic constraints on the column density of helium in the $2^{1}S$ state are lower limits, meaning that {\sc cloudy}-derived values must exceed the observed value (1.76$\times$10$^{12}$~cm$^{-2}$ from Table~\ref{tab:dintcols}) to be considered valid.  

The {\sc cloudy} output provides the fraction of helium in the sum of all singlet states at each principal quantum state rather than the individual singlet states at different total azimuthal quantum numbers. Helium in the n=2 singlet state exists in both $^{1}S$ and $^{1}P$ configurations, meaning the latter of these must be taken into account to derive the fraction in the $2^{1}S$ state. This state is formed by the combined effect of He$^{+}$ + e$^{-}$ recombinations and collisional excitation from the $2^{3}S$ metastable state. Approximately one third of helium recombinations to a singlet state of n=2 or higher ultimately results in population of the $2^{1}S$ state, while the rest lead to the ground state. Therefore, using the recombination and collisional rate coefficients derived from \citet{smits96,benjamin99} and \citet{bray00} respectively as listed in \citet{osterbrock06}, the fraction of helium in the $2^{1}S$ state at each {\sc cloudy} output zone can be found as follows

\begin{equation}
\label{eqn:he1spop}
F\left(2^{1}S\right)=F_{s}\times{}\frac{\alpha{}_{c1}n_{He}\left(2^{3}S\right)+0.33\alpha{}_{rec}n_{He+}}{\alpha{}_{c1}n_{He}\left(2^{3}S\right)+\alpha{}_{c2}n_{He}\left(2^{3}S\right)+\alpha{}_{rec}n_{He+}}\, ,
\end{equation}

\noindent where, for a specific zone, $F(2^{1}S)$ is the fraction of helium in the $2^{1}S$ state, $F_{s}$ is the {\sc cloudy}-calculated value of the helium fraction in the n=2 singlet states, $\alpha$$_{c1}$ is the $2^{3}S$--$2^{1}S$ collisional rate coefficient, $\alpha$$_{c2}$ is the $2^{3}S$--$2^{1}P$ collisional rate coefficient, $n_{{\rm He}}(2^{3}S)$ is the number density of helium in the $2^{3}S$ state, $n_{{\rm He}}(2^{1}S)$ is the number density of helium in the $2^{1}S$ state and $\alpha$$_{rec}$ is the sum of the recombination rate coefficients to all singlet states at n=2 or above. The bulk of the zone temperatures are within the ranges listed in tables 2.4 and 2.5 of \citet{osterbrock06}, therefore linear interpolation is used to find the appropriate rate coefficients at individual zones. For those zone temperatures outside the ranges listed, linear extrapolation is used.

\subsection{First coarse run}

It was convenient to initially generate a ``coarse'' grid measuring 8$\times$7 using a range in ionization parameter $U$ of $-$2.5$\leq$log$U$$\leq$1.0 in steps of 0.5 dex and a range in hydrogen number density $n_{{\rm H}}$ of 3.0$\leq$log($n_{{\rm H}}$/cm$^{-3}$)$\leq$9.0 in steps of 1.0 dex. Having found the gridpoints which satisfy the conditions derived from the spectra, a ``fine'' grid could then be calculated to narrow down the allowed values in the parameter space neighbouring the allowed coarse gridpoints. The gas hydrogen column density at each gridpoint was set equal to log$N_{{\rm H}}$=24.5+log$U$ to allow the code to integrate through the relatively high-ionization region occupied by $2^{3}S$ He~{\sc i}. The grid was generated by performing the following steps: (i) The He~{\sc i} $2^{3}S$-state column density output from {\sc cloudy} is compared to the epoch~5 spectroscopically-derived value. Gridpoints which produce values lower than this are ruled out, (ii) for those gridpoints that satisfy condition (i), the $2^{3}S$ He~{\sc i} number density in each output zone generated by the code is calculated by multiplying the fraction of helium in this state by the helium abundance and $n_{{\rm H}}$, (iii) the $2^{1}S$ He~{\sc i} number density in each output zone is found by calculating $F(2^{1}S)$ in each zone using Equation~\ref{eqn:he1spop} and multiplying the result by the helium abundance and $n_{{\rm H}}$, (iv) the absorber thickness is found by performing the integral $\int$$n_{{\rm He}}(2^{3}S)$dr through each zone from the irradiated face of the cloud inward until the epoch~5 column is reached, with the radius at which this value is reached giving the total thickness, (v) this integration is performed over the same zones as in (iv) to find the $2^{1}S$ column density using $\int$$n_{{\rm He}}(2^{1}S)$dr, and (vi) the total thickness is multiplied by $n_{{\rm H}}$ for the gridpoint in question to find the hydrogen column density $N_{{\rm H}}$.

\subsection{Second coarse run}

The absence of H$\beta$ absorption can be used to obtain an upper limit on the column density of hydrogen in the n=2 state and therefore rule out gridpoints where the {\sc cloudy} models suggest this upper limit is exceeded. To do this, {\sc cloudy} is run for a second time over the coarse grid, using a method very similar to that described in paragraph 3 of \S{}3.1 in \citet{leighly11}. The normalised epoch~5 profile shape for the He~{\sc i} $\lambda$3889 absorber is assumed to be the shape that any hydrogen absorption would follow. Balmer absorption can arise from hydrogen atoms in either of the two n=2 states ($^{2}S$ or $^{2}P$), so to calculate the correct column density the relative populations of atoms in each state must be known. Using the $N_{{\rm H}}$ values obtained from the first run, {\sc cloudy} was set to integrate through to these values and then run for a second time to determine the hydrogen n=2 population values. 

\begin{figure}
\centering
\resizebox{\hsize}{!}{\includegraphics[angle=0,width=8cm]{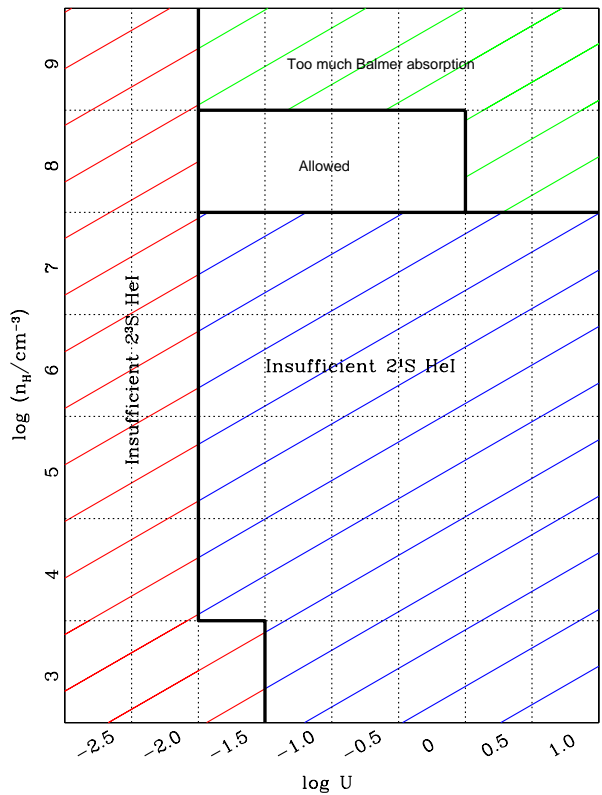}}
\caption{Allowed and ruled out coarse gridpoints (gridpoints are represented by rectangles enclosed by dotted lines). Gridpoints showing insufficient metastable $2^{3}S$ helium absorption were ruled out first (red shaded area). Gridpoints showing too much Balmer absorption were ruled out second (green shaded area). Gridpoints showing insufficient helium in the $2^{1}S$ state were ruled out third (blue shaded area). The remaining allowed region is unshaded.}
\label{fig:coarse}
\end{figure}

The minimum H~{\sc i} n=2 column density at which a measurement would become significant is estimated by replacing the f$\lambda$ value in Equation~\ref{eqn:ppc3} by a weighted mean value which depends upon the proportion of atoms in each of the two hydrogen n=2 states and using the same $C(v)$ profile as was calculated for $2^{3}S$ He~{\sc i}. The depth at every velocity bin of this profile is then scaled by factors of between 0.001 and 1 in steps of 0.001, with the column density calculated at each step using Equation~\ref{eqn:ppc3}. The first value of the column density at which the 1$\sigma$ uncertainty in the calculation is exceeded is assumed to be the maximum column density of hydrogen in the n=2 state. The constraints on the allowed coarse gridpoints obtained from the metastable Helium, n=2 singlet-state helium and maximum Balmer absorption are illustrated in Fig.~\ref{fig:coarse}

\subsection{Fine grid simulation}

A second set of {\sc cloudy} simulations using the same two-run method described in \S{}4.2 and \S{}4.3 was undertaken to further constrain the absorber properties at a finer resolution. A new grid measuring 24$\times$19 spanning the ranges $-$1.9$\leq$log$U$$\leq$0.4 and 7.1$\leq$log($n_{{\rm H}}$/cm$^{-3}$)$\leq$8.9, both in step sizes of 0.1 dex, was generated. Given the minimum value of $N_{{\rm limHe}}(2^{1}S)$=1.76$\times$10$^{12}$~cm$^{-2}$ listed in Table~\ref{tab:dintcols}, all log$n_{{\rm H}}$ values in the find grid are allowed. No attempt is made to rule out gridpoints in log$U$ space where the minimum singlet helium column density is not reached as the $2^{1}S$ He~{\sc i} column density is a weak function of log$U$ at a given log$n_{{\rm H}}$ in the grid. Further use is made of the fine grid in the Discussion section (\S{}5).

The presence of turbulence or differential velocity in the outflowing gas can reduce the Ly$\alpha$ optical depth and therefore lead to a higher gas density limit than that derived in these {\sc cloudy} models. This is due to reduced pumping into the H~{\sc i} n=2 state. Given the FWHM of the absorber ($\sim$600~km~s$^{-1}$) and the spectral resolution ($\sim$400~km~s$^{-1}$) the turbulence or differential velocity of the line cannot be larger than $\sim$500~km~s$^{-1}$. However it is likely to be lower than this, given the saturated nature of the near-IR line \citep{arav04}. Figure~22 (right panel) in Appendix~C of \citet{leighly11} shows that, for a $2^{3}S$ He~{\sc i} column density of the same order-of-magnitude, the upper limit on log($n_{{\rm H}}$/cm$^{-3}$) increases by about 1.5~dex at a turbulent velocity $v_{turb}$=500~km~s$^{-1}$. At $v_{turb}$$<$500~km~s$^{-1}$, the upper limit on log$n_{H}$ rapidly declines with decreasing $v_{turb}$.

\section{Discussion}

\subsection{Variability mechanism}

In \S{}4, the calculations were performed exclusively for epoch~5 due to the relatively narrow constraints obtained on the column density for that absorber, made possible by the detection of metastable helium absorption in both the optical (3889 \AA{}) and near-IR (10\,830 \AA{}) wavebands. All other epochs only detected a near-IR absorber, resulting in the calculation of lower column density limits only. Although absorption variability cannot be detected in the optical waveband, there is significant variability in the near-IR profile as indicated in Table \ref{tab:dintcols}. There are two main possibilities for this variability mechanism, either (a) changes in the ionization state of the gas under the influence of a varying incident ionizing continuum, or (b) the fraction of the emission source covered by the absorber changes over time due to the absorber's motion across the line-of-sight. These possibilities are now examined in detail. We only consider velocity space occupied by the epoch~5 absorber due to the availability of the simulation results and the lack of constraints on the (possibly physically separate) absorber at higher outflow (more negative) velocities.

\subsection{Are ionizing continuum changes responsible?}

Many AGN are known to have variable continuum emission in the optical and UV wavebands. This is true of NGC 4151 which shows variability across X-ray, UV and optical wavebands on timescales ranging from hours/days to multiple years e.g. \citet{edelson96,crenshaw96,kong06}. Ionization of gas in the AGN vicinity is mainly controlled by the incident extreme ultraviolet (EUV) flux originating from the continuum emission region. If the gas is sufficiently dense, resulting in short recombination timescales, changes in the strength of associated emission/absorption lines should be correlated with changes in the unobserved EUV continuum. It can be assumed that the optical-UV power-law continuum extends into the EUV, meaning measurements of the behaviour of the optical continuum can be used as a proxy for (ionizing) EUV variations \citep{kriss99}. This is the basis of reverberation mapping, where it is known that changes in the broad emission lines correlate with changes in the optical continuum, e.g. \citet{peterson98}. However, this correlation may not be true for all AGN absorption lines. Studies showing evidence for ionizing continuum driven variability in AGN absorption lines include \citet{wildy15,wang15}.

To test the plausibility of ionization-driven variability, we measured the relative changes in the flux of the power-law continuum model. The photometric accuracy of the spectra is not sufficient to directly compare continuum strengths across the observational epochs. Instead we scale the flux using the forbidden narrow line [Ne~{\sc iii}] $\lambda$3868, which is not expected to vary between epochs due to the spatially extended nature of the NLR and the low responsivity of forbidden lines to ionizing flux changes. The following procedure is therefore undertaken. At each epoch we multiply the power-law model by the flux of the [Ne~{\sc iii}] $\lambda$3868 template before dividing by the flux of the epoch 6 template and finally dividing the result by the power-law model at epoch 6 (template construction was described in \S{}3.2.3). As noted, no template was generated for epoch~1, so instead at this epoch we use the same flux ratio but over only the unblended (blue side) of the line after subtracting a local linear continuum. The results are flux-corrected continuum models normalised to epoch~6. The values recorded at 4000~\AA{} for each epoch are displayed in Table~\ref{tab:cgratios}. The continuum peak--trough periodicity is consistent with V-band photometry of NGC~4151 recorded in \citet{guo14}, which overlaps our epochs~3--5. 

\begin{table}
\begin{center}
\caption{Epoch 6-normalised power-law continuum model fluxes at 4000~\AA{}.}
\begin{tabular}{l l}
\hline Epoch&Normalised flux\\
 & \\
\hline
1&5.1$\pm$0.8\\
2&2.5$\pm$1.8\\
3&1.7$\pm$1.3\\
4&1.0$\pm$0.6\\
5&9.9$\pm$0.8\\
6&1.0$^{\dagger}$$\pm$0.5\\
\hline
\end{tabular}
\label{tab:cgratios}
\end{center}
$^{\dagger}$ value is unity by definition
\end{table}

\begin{figure}
\centering
\resizebox{\hsize}{!}{\includegraphics[angle=0,width=8cm]{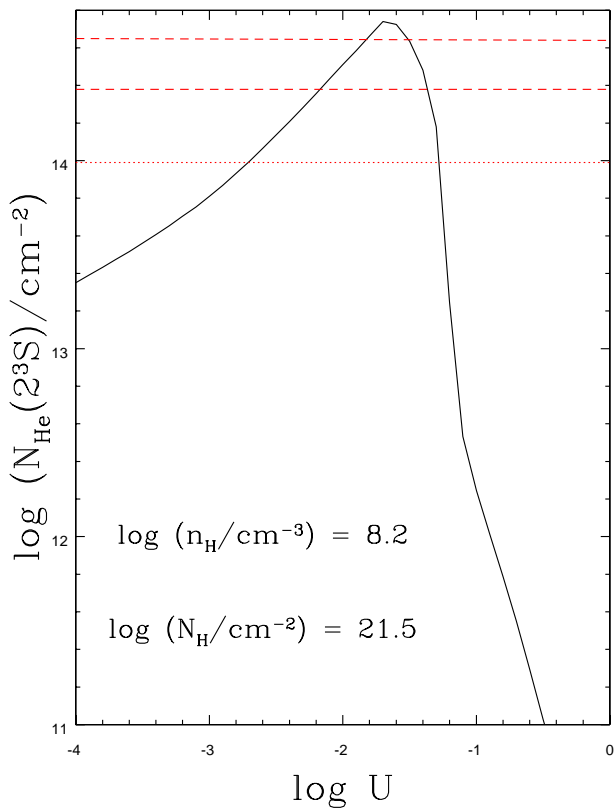}}
\caption{Behaviour of the $2^{3}S$ helium column density as a function of ionization parameter at a fixed hydrogen parameters of log($n_{{\rm H}}$/cm$^{-3}$)=8.2, log($N_{{\rm H}}$/cm$^{-2}$)=21.5. The epoch~5 (PPC model) absorber 1$\sigma$ upper and lower column density limits are indicated by the dashed red lines. The dotted red line indicates the corresponding upper limit allowed by scaling the epoch~5 near-IR profile to the epoch~1 detection limit.}
\label{fig:uvar}
\end{figure}

If continuum changes are driving the variability, it is important to understand in what sense the absorber should respond. For the epoch~5 absorber, Fig.~\ref{fig:uvar} shows a representative curve of absorber column density varying as a function of log$U$ taken at example physical conditions allowed by the {\sc cloudy} fine-grid simulation (log($n_{{\rm H}}$/cm$^{-3}$)=8.2 cm$^{-3}$, log($N_{{\rm H}}$/cm$^{-2}$)=21.5). Since log$n_{{\rm H}}$ is held constant, variations in log$U$ must correspond to changes in ionizing photon flux. From epoch~1, the continuum drops from a high state to a low trough at epochs~3 and 4, before reaching to its highest value of all 6 observations at epoch~5. It then returns to a lower state at epoch~6. Given the shape of the column density response in Fig.~\ref{fig:uvar}, this pattern of continuum variability would have to explain the absence of the epoch~5 absorber at epochs 1 and 2 (possibly also absent at epochs 3 and 4) and its subsequent disappearance at epoch~6. This is consistent with a situation where log$U$ is initially sufficiently low that any $2^{3}S$ He~{\sc i} absorption is undetectable (left of the peak in Fig.~\ref{fig:uvar}). The subsequent highly significant increase in continuum flux at epoch~5 boosts the metastable hydrogen column density to within the limits allowed by the partial coverage model, with the situation reversing between epochs~5 and 6.

From the example in Fig.~\ref{fig:uvar}, a reduction in log$U$ of at least $\sim$0.6~dex is required to remove the epoch~5 absorber from detectability at epochs~1 and 6. Given the fact that variability at ionizing wavebands can be many times greater than that in the optical e.g. \citet{edelson96}, this is achievable, as the 4000~\AA{} flux ratios at these epochs relative to epoch 5 are $\sim$0.5 and $\sim$0.1 respectively. The observations therefore support a situation where changes in the ionizing continuum incident on the absorber generate its measured variability.

\subsection{A crossing-clouds model}

The plausibility of changes in gas covering fraction as an explanation for the observed variability can be tested using a method similar to the ``crossing disc'' model applied in \citet{capellupo13,capellupo14}. In this model a disc-shaped absorber moves across a disc-shaped emission source, changing the fraction of the emission covered and hence the apparent strength of the absorber. Given the values for the near-IR absorber recorded in Table~\ref{tab:dintcols}, a plausible scenario that occurred in NGC 4151 over the course of the observations is that of a cloud starting to eclipse the AGN emission region as viewed from Earth at some point between epochs 2 and 3, reaching a peak coverage near epoch 5 before moving out of the line-of-sight prior to the epoch~6 observation. For the purposes of this study the model relates the velocity of gas across the line-of-sight to the crossing time as follows

\begin{equation}
\label{eqn:crossdisc}
v_{c}=\frac{\sqrt{\Delta{}A}D_{{\rm c}}}{\Delta{}t}\, ,
\end{equation}

\noindent where $v_{c}$ is the velocity of a disc-shaped cloud at right angles to the line-of-sight, $\Delta$A is the change in the covering fraction of the emission source, $D_{{\rm c}}$ is the diameter of the region emitting the power-law continuum and $\Delta$t is the cloud crossing time available in the AGN rest-frame during the epoch interval. Assuming an optically thick, geometrically thin accretion disc \citep{shakura73}, an annulus at a given radial distance from the SMBH emits as a blackbody, with temperature dependent on distance. Therefore an estimate can be made of the size of the continuum region at the absorber centroid by considering all emission to be within a radius defined by the temperature giving the peak flux at this wavelength. This provides a value of $D_{{\rm c}}$=1.1$\times$10$^{16}$~cm. Covering fractions are calculated using the average normalised flux of the near-IR absorption component in the spectral range spanned by the epoch~5 absorber. These are listed in Table~\ref{tab:cfvals}.

\begin{table}
\begin{center}
\caption{Fraction of emission region covered by absorber at each epoch}
\begin{tabular}{l l}
\hline Epoch&Covering Fraction ($A$)\\
\hline
1&0$\pm$0.09\\
2&0$\pm$0.14\\
3&0.16$\pm$0.06\\
4&0.10$\pm$0.14\\
5&0.38$\pm$0.03\\
6&0$\pm$0.13\\
\hline
\end{tabular}
\label{tab:cfvals}
\end{center}
\end{table}

Covering fraction changes ($\Delta$A) are computed across time intervals beginning or ending with epoch~5. Epoch~1 is excluded as the sense of any change between epochs~1 and 2 is completely unknown. Therefore, calculations for the intervals between epochs 2--5, 3--5, 4--5 and 5--6 are performed. Checking the consistency of the crossing-clouds model requires calculating the value of $\sqrt{\Delta{}A}$/$\Delta$t and its associated error at each epoch interval used. These are listed in Table~\ref{tab:deltaA}.

\begin{table}
\begin{center}
\caption{Epoch interval-time-normalised $\sqrt{\Delta{}A}$/$\Delta$t values}
\begin{tabular}{l l}
\hline Epoch interval&$\sqrt{\Delta{}A}$/$\Delta$t\\
 &(10$^{-9}$~sec$^{-1}$)\\
\hline
2--5&3.38$\pm$0.67\\
3--5&3.94$\pm$0.66\\
4--5&5.37$\pm$1.40\\
5--6&4.28$\pm$0.63\\
\hline
\end{tabular}
\label{tab:deltaA}
\end{center}
\end{table}

The velocity values calculated at the two epoch intervals showing significant absorption at both endpoints (3--5 and 4--5) are 420~km~s$^{-1}$ and 570~km~s$^{-1}$ respectively. The best value is therefore taken as the mean of these ($v_{c}$=495~km~s$^{-1}$) under the assumption that crossing-clouds model is correct. One possible explanation for movement of absorbing gas across the line-of-sight to an AGN is that the gas is in orbit around the black hole with a lateral Keplerian velocity. The radial distance in such a scenario is $r_{{\rm kep}}$=GM/$v_{c}$$^{2}$. Using the estimated mass of the black hole in NGC 4151 of $M_{{\rm BH}}$=4.57$\times$10$^{7}$ $M_{\odot}$ from \citet{bentz06b}, a value of $r_{{\rm kep}}$=2.5$\times$10$^{18}$~cm (0.81~pc) is obtained.

If a crossing-clouds model were correct, consistency would be expected between values of $\sqrt{\Delta{}A}$/$\Delta$t recorded at intervals 3--5 and 4--5. Additionally, intervals 2--5 and 5--6 should each show values consistent with or less than those at intervals 3--5 and 4--5, since the time the absorber spends outside the line-of-sight between epochs~2 and 3 and epochs~5 and 6 is unknown. From Table~\ref{tab:deltaA} it is clear that these conditions are met. Although this leads to self-consistency within the crossing clouds scenario, it does not provide conclusive evidence of its reality. Such evidence would require: (a) tightly constrained non-zero covering fractions across several epochs, and (b) knowledge that the same absorber is likely recorded at each epoch. It also seems coincidental within this model that the peak covering fraction should occur at maximum continuum strength. Therefore ionization change driven by the continuum is favoured as the variability mechanism.

\subsection{Allowed parameter space and mass outflow rates of the absorber}

\begin{figure}
\centering
\resizebox{\hsize}{!}{\includegraphics[angle=0,width=8cm]{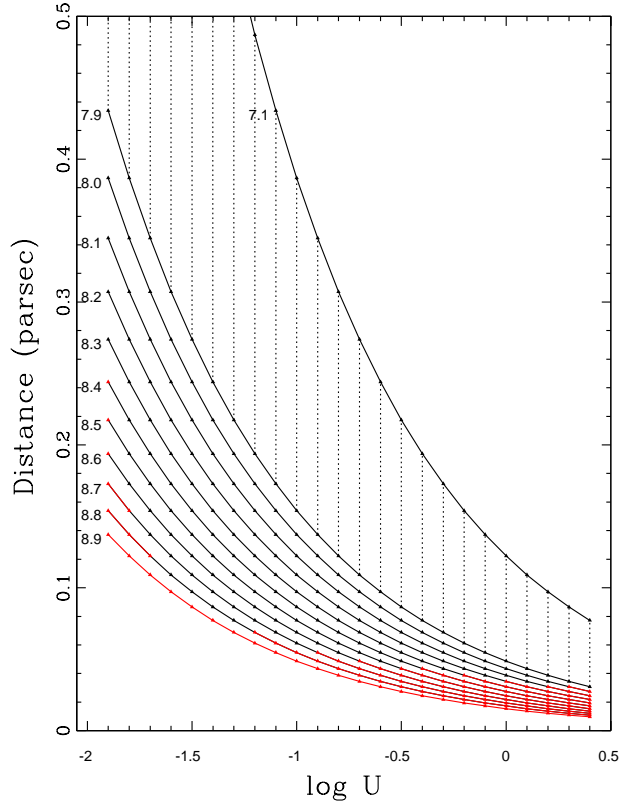}}
\caption{Radial location of the absorber as a function log$U$ at log$n_{{\rm H}}$ values (indicated by numbers left of the respective curves). Fine gridpoints are indicated by triangles. Regions on the curve allowed by the Balmer absorption criterion are in black, regions predicting too much Balmer absorption are in red. All log$U$ values in the fine grid are allowed at 7.1$\leq$log$n_{{\rm H}}$$\leq$7.9 (region indicated by vertical dotted lines). Maximum distance (off-scale at log$U$=$-$1.9, log$n_{{\rm H}}$=7.1) is 1.1~pc.}
\label{fig:distvals}
\end{figure}

The un-normalised SED used in the {\sc cloudy} calculations can be appropriately scaled using the value of the bolometric luminosity given by $L_{{\rm bol}}$=7.3$\times$10$^{43}$ erg s$^{-1}$ \citep{kaspi05}. Using this scaled SED, we can estimate the distance from the ionizing continuum source to the incident surface of the epoch~5 absorber at each of the fine gridpoints described in \S{}4.4, as shown in Fig.~\ref{fig:distvals}. Radial distances are within 0.49~pc of the continuum source for 90 per cent of all valid gridpoints, with an inner limit of 0.03~pc.

\begin{figure}
\centering
\resizebox{\hsize}{!}{\includegraphics[angle=0,width=8cm]{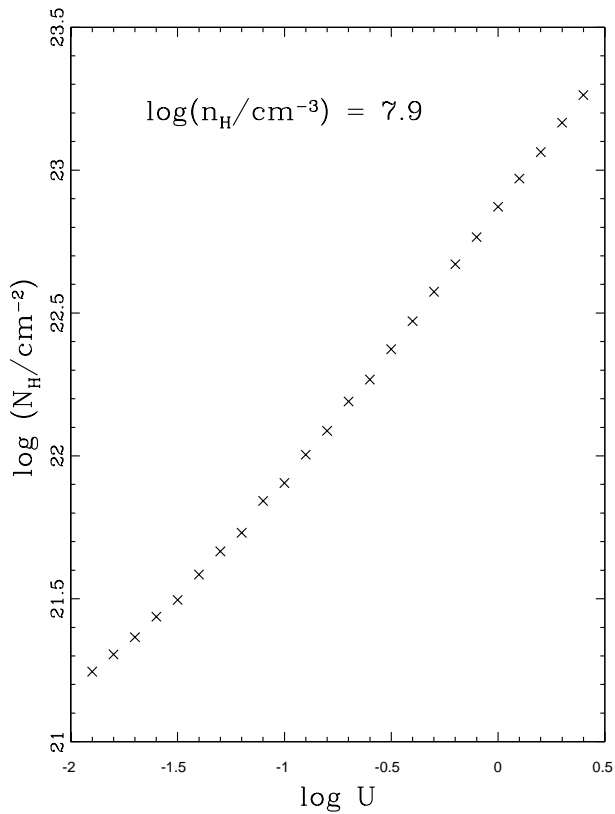}}
\caption{Plot of log$N_{{\rm H}}$ against log$U$ at log($n_{{\rm H}}$/cm$^{-3}$)=7.9. Gridpoints are indicated by black crosses.}
\label{fig:hcdens}
\end{figure}

At a given log$U$ gridpoint, total hydrogen column densities are a very weak function of log$n_{{\rm H}}$. Therefore only one set of points corresponding to log($n_{{\rm H}}$/cm$^{-3}$)=7.9, representing log$N_{{\rm H}}$ as a function of log$U$, is indicated in Fig.~\ref{fig:hcdens}. Allowed limits are 21.2$\leq$log($N_{{\rm H}}$/cm$^{-2}$)$\leq$23.3. Hydrogen column densities are required to calculate mass outflow rates, can be estimated from total hydrogen column densities using the method of \citet{crenshaw09} as follows

\begin{equation}
\label{eqn:mout}
\dot M_{{\rm out}}=8\pi{}RN_{{\rm H}}\mu{}m_{{\rm p}}C_{{\rm AGN}}|v_{m}|\, ,
\end{equation}

\noindent where $\mu$ is the mean atomic mass per proton ($\mu$$\sim$1.4), $m_{{\rm p}}$ is the proton mass, $C_{{\rm AGN}}$ is the covering fraction of the outflow as seen from the active nucleus and $v_{m}$ is the mean velocity of the outflow. Using an 80$^{\circ}$ opening angle \citep{hutchings98}, the covering fraction can be estimated as $C_{{\rm AGN}}$=0.23. The velocity of the (predominantely hydrogen) outflow is assumed to be approximately the same as the velocity centroid of the helium absorption ($v_{m}$$\sim$400~km~s$^{-1}$). Examination of the outflow's hydrogen column density, radial distance and outflow rate across the permitted gridpoints allows constraints to be placed on these quantities. For the mass outflow rate the limits are 0.008$\leq$$M_{{\rm out}}$~($M_{\odot}$~yr$^{-1}$)$\leq$0.38, which corresponds to a kinetic luminosity range of 4.11$\times$10$^{38}$ to 1.90$\times$10$^{40}$~erg~s$^{-1}$. The weakness of this outflow, with a kinetic luminosity of at most 0.026 per cent of the bolometric luminosity, indicates that it is not a significant contributor to galaxy feedback \citep{hopkins10}. The derived range in outflow rate overlies the relatively low mass accretion rate in NGC 4151 of $\dot M_{acc}$=0.013~$M_{\odot}$~yr$^{-1}$ as calculated by \citet{crenshaw07}. At its lowest possible value, the outflow is about $\sim$62 per cent of the accretion rate, however at its maximum it is $\sim$29 times larger.

\subsection{Comparison with previous observations}

Since cases of absorption from metastable triplet state He~{\sc i} are effectively high ionization transitions, the observed absorption may also be visible in high ionization UV transitions resulting from species such as C~{\sc iv} and P~{\sc v}. Although no simultaneous UV spectra are available, the centroid of the epoch~5 absorber is similar to the 'D+E' feature described in \citet{kraemer05,kraemer06} and \citet{crenshaw07}. The values listed in \citet{crenshaw12} for the ionization parameter (log$U$=$-$1.08), radial distance ($\sim$0.1~pc) and mass outflow rate (0.014~$M_{\odot}$~yr$^{-1}$) found for the UV sub-component labelled 'D+Ec' lie within the range allowed in Fig.~\ref{fig:distvals}. Therefore the $2^{3}S$ state-helium absorber studied in this paper is consistent with the physical properties of 'D+E' without the high cloumn density components labelled 'D+Ea' and 'Xhigh' contributing. The lack of high column absorbers in our study may not be totally surprising given the lack of detection of H$\beta$ absorption, in contrast to that found in the June 2000 HST observation reported in \citet{crenshaw07}. It is likely that our epoch~5 observations detect an outflow between the inner narrow line region and the BLR, possibly within an \emph{intermediate line region} where self-absorption occurs \citep{crenshaw07}. A hydrogen number density range of 7.1$\leq$log($n_{{\rm H}}$/cm$^{-3}$)$\leq$8.8 is intermediate to the BLR and NLR densities and is therefore consistent with the location found in the aforementioned papers. If this UV absorption results from the same obscurer as our epoch~5 observation, it rules out the crossing cloud scenario since, at the velocities listed in \S{}5.3, the absorber would have completely crossed the emission region at least $\sim$3 years before epoch~5.

Given the range of possible radial distances (see Fig.~\ref{fig:distvals}), the $2^{3}S$ helium epoch~5 absorber likely lies at or beyond the radial distance of the inner edge of the dusty torus. The torus is thought to surround the central regions of AGN including NGC 4151, with a structure extending from the inner face out to a few pc \citep{riffel09}. In \citet{koshida14} the observed time lag between the V-band and K-band light-curve behaviour in NGC 4151 suggested an inner radius of the dust torus of approximately 30 to 80 lt-days. The earlier study of \citet{minezaki04} was consistent with this finding, having calculated a hot inner edge of the torus in NGC 4151 located at $\approx$0.04 pc from the central ionising source. It is therefore possible that the outflow consists of material originating in the torus, as has been proposed for a broad absorption line in \citet{leighly15}.

\section{Summary}

The results of this investigation can be summarised as follows:\\

\noindent (i) The absorption due to He~{\sc i} $\lambda$20\,587 has a similar profile to that at 10\,830~\AA{}, indicating that they originate in the same outflowing gas. Use of this line, which is not commonly examined in AGN studies, in combination with the triplet state transitions, provides useful constraints on the parameter space occupied by the outflow.

\noindent (ii) The near-IR component of the triplet state helium absorber in epoch~5 is likely to be saturated over most of its velocity range, since the depth of the optical component is significantly stronger than that predicted by an absorption profile obtained by scaling down the optical depth of the near-IR component by a factor 23.3 based on the f$\lambda$ ratio.

\noindent (iii) The continuum behaviour, together with the variability pattern across the observational epochs, suggests changes in the ionization state driven by a variable continuum are the cause of the observed variability of the epoch~5 absorber.

\noindent (iv) The observed variability is consistent with changes in absorber covering fraction of the emission region. The absorber strength is consistent with zero, therefore being entirely outside the line-of-sight, at epochs 1 and 2. It would reach peak coverage near epoch~5 before moving out again prior to epoch~6. However, the lack of multiple significant absorber measurements precludes conclusive evidence of this mechanism.   

\noindent (v) Photoionization simulations are used to constrain the epoch~5 absorber properties. The total hydrogen number density and column density ranges are 7.1$\leq$log($n_{{\rm H}}$/cm$^{-3}$)$\leq$8.8 and 21.2$\leq$log($N_{{\rm H}}$/cm$^{-2}$)$\leq$23.3 respectively, while the ionization parameter range is $-$1.9$\leq$log$U$$\leq$0.4. Ninety per cent of the gridpoints record radial distance from the continuum source of 0.03$\leq$r$\leq$0.49~pc with a maximum permitted value of 1.1~pc.

\noindent (vi) The epoch~5 absorber shows similarities to that arising from the absorber complex labelled 'D+E' in NGC 4151 reported in \citep{kraemer05,kraemer06,crenshaw07,crenshaw12}. The properties of the epoch 5 absorber are what would be expected if the highest column density components of 'D+E' were excluded. The lack of H$\beta$ absorption in our spectra suggests these high column components are indeed no longer present.

\noindent (vii) The mass outflow rate due to the epoch~5 absorber is calculated to be between 0.008 and 0.38~$M_{\odot}$ yr$^{-1}$. This outflow rate implies a contribution to the kinetic luminosity which is low in comparison to the bolometric luminosity, therefore the outflow is not a significant contributor to feedback effects. Compared to the mass accretion rate, the possible mass outflow rate ranges from approximately 0.6 to 29 times as large.

\section*{Acknowledgements}

We thank the anonymous referee for their helpful comments which led to the improvement of this paper. This work is supported at the University of Leicester by the Science and Technology Facilities Council (STFC). HL is supported by a European Union COFUND/Durham Junior Research Fellowship (under EU grant agreement number 267209).

\bibliography{bib_cw}

\bibliographystyle{mn2e}

\bsp

\label{lastpage}

\end{document}